\documentclass[final,3p,times,twocolumn]{elsarticle}

\usepackage{tcolorbox}

\newcommand{\finding}[1]{
\begin{tcolorbox}[leftrule=1mm,toprule=0mm,bottomrule=0mm,left=1pt,right=2pt,top=2pt,bottom=2pt]
\em #1
\end{tcolorbox}
}

\newcommand{\rqone}{What are the main software engineering areas and tasks for which Explainable AI approaches have been used to date?}
\newcommand{\rqoneOne}{What are the main software engineering areas and tasks for which XAI has been applied?}
\newcommand{\rqonetwo} {What are the ML4SE works that offer explainability?}
\newcommand{\rqonethree} {What are the objectives of explainability in each research paper reviewed?}
\newcommand{\rqtwo}{What are the Explainable AI approaches adopted for SE tasks?}
\newcommand{\rqtwoone} {What kind of XAI techniques have been used?}
\newcommand{\rqtwotwo} {What kind of explanations do they offer?}
\newcommand{\rqthree}{How useful XAI has been for SE?}
\newcommand{\rqthreeone} {What are the objectives of applying XAI on SE and how they have been useful?}
\newcommand{\rqthreetwo} {What are the means of evaluating XAI in SE?}

\usepackage{amssymb}
\usepackage{longtable}
\usepackage{multirow}
\usepackage[justification=centering]{caption}% or e.g. [format=hang]
\journal{Information and Software Technology}

\begin{document}

\begin{frontmatter}

%% Title, authors and addresses

%% use the tnoteref command within \title for footnotes;
%% use the tnotetext command for theassociated footnote;
%% use the fnref command within \author or \address for footnotes;
%% use the fntext command for theassociated footnote;
%% use the corref command within \author for corresponding author footnotes;
%% use the cortext command for theassociated footnote;
%% use the ead command for the email address,
%% and the form \ead[url] for the home page:
%% \title{Title\tnoteref{label1}}
%% \tnotetext[label1]{}
%% \author{Name\corref{cor1}\fnref{label2}}
%% \ead{email address}
%% \ead[url]{home page}
%% \fntext[label2]{}
%% \cortext[cor1]{}
%% \affiliation{organization={},
%%             addressline={},
%%             city={},
%%             postcode={},
%%             state={},
%%             country={}}
%% \fntext[label3]{}

\title{A Systematic Literature Review of Explainable AI for Software Engineering}

%% use optional labels to link authors explicitly to addresses:
%% \author[label1,label2]{}
%% \affiliation[label1]{organization={},
%%             addressline={},
%%             city={},
%%             postcode={},
%%             state={},
%%             country={}}
%%
%% \affiliation[label2]{organization={},
%%             addressline={},
%%             city={},
%%             postcode={},
%%             state={},
%%             country={}}

\author[inst1]{Ahmad Haji Mohammadkhani}

\affiliation[inst1]{organization={University of Calgary},%Department and Organization
            % addressline={}, 
            city={Calgary},
            % postcode={00000}, 
            state={Alberta},
            country={Canada}}
            
\author[inst2]{Nitin Sai Bommi}

\affiliation[inst2]{organization={University of Hyderabad},%Department and Organization
            % addressline={}, 
            city={Hyderabad},
            country={India}}

\author[inst3]{Mariem Daboussi}

\affiliation[inst3]{organization={INSAT },%Department and Organization
            % addressline={}, 
            city={Tunis},
            country={Tunisia}}
            
\author[inst4]{Onkar Sabnis}

\affiliation[inst4]{organization={Indian Institute of Technology Kharagpur},%Department and Organization
            % addressline={}, 
            city={Kharagpur, West Bengal},
            country={India}}

\author[inst5]{Chakkrit Tantithamthavorn}
% \author[inst1,inst2]{Author Three}
\affiliation[inst5]{organization={Monash University},%Department and Organization
            % addressline={Address Two}, 
            city={Melbourne},
            % postcode={22222}, 
            state={Victoria},
            country={Australia}}

\author[inst6]{Hadi Hemmati}
% \author[inst1,inst2]{Author Three}
\affiliation[inst6]{organization={York University},%Department and Organization
            % addressline={Address Two}, 
            city={Toronto},
            % postcode={22222}, 
            state={Ontario},
            country={Canada}}

\begin{abstract}
%% Text of abstract
\textbf{Context:} In recent years, leveraging machine learning (ML) techniques has become one of the main solutions to tackle many software engineering (SE) tasks, in research studies (ML4SE). This has been achieved by utilizing state-of-the-art models that tend to be more complex and black-box, which is led to less explainable solutions that reduce trust and uptake of ML4SE solutions by professionals in the industry.

\textbf{Objective:} One potential remedy is to offer explainable AI (XAI) methods to provide the missing explainability. In this paper, we aim to explore to what extent XAI has been studied in the SE community (XAI4SE) and provide a comprehensive view of the current state-of-the-art as well as challenge and roadmap for future work. 

\textbf{Method:} We conduct a systematic literature review on 24 (out of 869 primary studies that were selected by keyword search) most relevant published studies in XAI4SE. We have three research questions that were answered by meta-analysis of the collected data per paper.

\textbf{Results:} Our study reveals that among the identified studies, software maintenance (\%68) and particularly defect prediction has the highest share on the SE stages and tasks being studied. Additionally, we found that XAI methods were mainly applied to classic ML models rather than more complex models. We also noticed a clear lack of standard evaluation metrics for XAI methods in the literature which has caused confusion among researchers and a lack of benchmarks for comparisons.

\textbf{Conclusions:} XAI has been identified as a helpful tool by most studies, which we cover in the systematic review. However, XAI4SE is a relatively new domain with a lot of untouched potentials, including the SE tasks to help with, the ML4SE methods to explain, and the types of explanations to offer. This study encourages the researchers to work on the identified challenges and roadmap reported in the paper.

\end{abstract}

% %%Graphical abstract
% \begin{graphicalabstract}
% \includegraphics{grabs}
% \end{graphicalabstract}

% %%Research highlights
% \begin{highlights}
% \item Research highlight 1
% \item Research highlight 2
% \end{highlights}

\begin{keyword}
%% keywords here, in the form: keyword \sep keyword
Explainable AI for Software Engineering (XAI4SE) \sep Systematic Review, \sep Machine Learning for Software Engineering (ML4SE) \sep Explainable AI \sep Interpretable AI
%% PACS codes here, in the form: \PACS code \sep code
% \PACS 0000 \sep 1111
%% MSC codes here, in the form: \MSC code \sep code
%% or \MSC[2008] code \sep code (2000 is the default)
% \MSC 0000 \sep 1111
\end{keyword}

\end{frontmatter}

%% \linenumbers

% \newpage
% \section*{Declarations:}
% \textbf{Funding}\\
% Haji Mohammadkhani and Hemmati  acknowledge the support of the Natural Sciences and Engineering Research Council of Canada (NSERC) and Alberta Innovates, [ALLRP/556396-2020 and ALLRP/568643-2021] and Tantithamthavorn acknowledge the support of the Australian Research Council's Discovery Early Career Researcher Award (DECRA) funding scheme (DE200100941). \\
% \textbf{Conflict of interest}\\
% The authors have no conflicts of interest to declare relevant to this article's content.\\
% \textbf{Data Transparency}\\
% All Data relevant to the experiments are available in the Github repository (or are taken from the public datasets).\\

% \newpage\newpage
%% main text
\newpage
\section{Introduction}

\label{intro}

In the past few decades, advancement in Machine Learning (ML) has exponentially increased with the combination of powerful machines, robust algorithms, and easier access to vast amounts of data. 
As a result, at present, ML models have been developed in many critical domains such as in healthcare, banking, finance, terrorism detection~\citep{doshi2014comorbidity,khandani2010consumer,le2018predicting}.

In the field of software engineering as well, ML has been dominating many studies and activities. AI-based solutions have been applied on many automation tasks such as source code~\citep{svyatkovskiy2020intellicode}, test case~\citep{koo2019pyse}, patch~\citep{white2019sorting}, specification generation~\citep{ataiefard2021deep}, prediction tasks such as defect or vulnerability prediction~\citep{dam2018automatic}, and recommendation tasks such as API recommendation~\citep{he2021pyart}. 

Easy access to an abundance of software repositories (e.g., GitHub, StackOverflow, execution logs, etc.) has made SE an ideal field for data-driven ML algorithms to thrive.
During the past decade, there has been many publications in sub-domains of SE research such as Mining Software Repositories (Software Analytics) and Automated Software Engineering that study applying ML techniques on various SE tasks such as API recommendation~\citep{he2021pyart}, risk prediction~\citep{filippetto2021risk}, code comprehension~\citep{ben2018neural}, code generation~\citep{svyatkovskiy2020intellicode}, code retrieval~\citep{gu2021cradle}, code summarization~\citep{shido2019automatic}, bug-fixing processes~\citep{tufano2019empirical}, software testing~\citep{koo2019pyse}, fault localization~\citep{miryeganeh2021globug}, effort estimation~\citep{phannachitta2020optimal}, documentation ~\citep{feng2020codebert}, clone detection~\citep{perez2019cross}, program synthesis~\citep{shin2019program}, etc.

Often case, the models with higher complexity such as SVMs~\citep{xu2018prediction}, and deep neural networks~\citep{mashhadi2020hybrid} have achieved higher predictive accuracy in these tasks and thus recommended to be used by practitioners. 
However, similar to other fields, like NLP or Image Processing, as the ML models become more complex and achieve better accuracies, understanding their decision-making process becomes much harder. State-of-the-art models evolved from using algorithms like decision trees or logistic regression that are intrinsically explainable, to using Deep Learning (DL) architectures, which basically offer no explanation for their decisions. Even though the final results may improve, more questions arise regarding their black-box nature. 

This black-box nature of the solution can cause issues with trusting these models at different levels. It will provoke distrust in managers and customers, especially in more critical tasks. For instance, in the case of SE tasks whenever an automated solution is suggested to be deployed, e.g., a generated code or patch, etc.) where the wrong predictions/generations may cause expensive failure or extra manual overhead to check and fix, leading to a lack of adoption in practice~\cite{dam2018explainable}. Also, ethical challenges and the risk of immoral biases emerge~\citep{chakraborty2019software} and it leaves the researchers with less clue about the right direction for improving their models~\citep{amershi2019software}. 

In other words, acquiring the explainable output from the so-called black-box models, is one of the key problems of the existing machine learning models, to be adopted in practice. Though the black-box models themselves learn the connection between the features and the final output, they do not clearly describe the actual relationship between the given set of features and the individual prediction bestowed by the models. It is extremely important to know why a model makes a specific prediction to retain the model's reliability, in many domain including most SE tasks. For that, the model should be able to offer explanations for an individual prediction or the model's overall approach to the practitioners.

%A similar problem holds when applying ML in SE tasks. 
Software practitioners as the main users of ML models in SE would only be interested in adopting ML-based code recommendation, bug repair tools, etc., if they are persuaded that the models have learned what they ``should have'' and predict or generate what they are ``meant to'', from the perspective of the human experts. Without establishing such trust any solution is seemed as unreliable with potential unexpected consequences that wastes more resources and time, than it has saved by automation. In addition, if the outputs of the models are not informative enough, they can only lead to more ambiguity rather than helping the developers and users.

As another example, let's look at the defect prediction problem, which is one of the most studied tasks in SE by researchers when it comes to applying ML in SE. It has taken advantage of ML/DL models for many years. The goal is to predict if a piece of code or file is defective or not. In this scenario, if the model finds a file defective, but not any more information about its decision, the developer will have no clue on what to look for if the model provides the reasons for the file's being defective or specify some lines or tokens as the possible defective parts, it would be easy for the software engineers to minimize the bugs based on the grounds provided and validate the ML-based recommendation.   

In short, model explainability facilitates the debugging process, bias detection (recourse to individuals who are adversely affected by the model predictions), assess when to trust model predictions in decision-making, and determining if they are suitable for deployment.

Thus, there is a compelling need to overcome the trade-off between accuracy and interpretability to provide trustworthy, fair, robust models for real-world applications. As a consequence, AI researchers have focused on the space of eXplainable Artificial Intelligence (XAI)~\citep{gunning2019darpa}, which provides an explainable and understandable interpretation of the behavior of AI systems.

Responding to the growing attraction in the industry and academia to use AI models in SE in recent years and the critical need for interpretation of AI models that are getting more complex, we conducted a systematic literature review (SLR). The SLR is based on relevant journal and conference papers that offer explainability methods for SE tasks or propose ML models that include any type of explanation.
We conducted this research to provide practitioners and researchers a better insight into the efforts that have been made in this topic so far and the state-of-the-art of XAI in software engineering. In order to do so we have defined three research questions. 

\begin{itemize}
    \item\textbf{RQ1: \rqone}
\end{itemize}
RQ1 aims to understand to what extent XAI has been used already in the SE community, which can potentially reveal which areas need further study. As our analysis show, ``software maintenance'' is the highest explored area and defect prediction is the most favorable task for the XAI researchers in SE by far, but areas such as testing and program repair have not used XAI much, even though they have used ML a lot.
\begin{itemize}
    \item\textbf{RQ2: \rqtwo}
\end{itemize}
The goal of RQ2 is to understand what XAI methods the SE community are using, which in turn will help us guide the future work by identifying methods that have shown good results already and potetnial methods that are not studied yet. 
Our analysis shows that most of the explanations so far are coming from self-explaining models and in the form of model internals (e.g., variable coefficients). Also, we can see that LIME and ANOVA are two of the mostly utilized XAI methods. We also see that higher-level explanations such as visualization and explanations in natural language are less sutied in XAI4SE.
\begin{itemize}
    \item\textbf{RQ3: \rqthree}
\end{itemize}
In RQ3, we focus on the usefulness of XAI methods and how the offered explanation has helped the users to improve or better understand the models. The goal is to have a realistic view if the potential benefits that can steer the future research in this domian. To answer this question, we collect the original authors' view, based on their results, on the usefulness, not our subjective opinion. One of the most interesting improvements reported in the studies is the usage of XAI to make the defect prediction models more precise while finding bugs and defects. Also, we discuss the lack of standard evaluation metrics according to the studies and how human-centered evaluations should be beneficial.

Finally, we have summarized the limitations and challenges of XAI4SE and provided a roadmap for future works.

To the extent of our knowledge, this is the first SLR for XAI in software engineering. The key contributions of this paper are:
\begin{itemize}
\item We present a consolidated body of knowledge of research on different aspects of XAI in software engineering 
\item  We conducted a comprehensive search on the most influential venues in the field and selected 24 related papers that match our criteria.
\item We analyzed these 24 papers to get a better insight into different characteristics of the problem in hand (e.g. their objective, datasets, representation, etc.), the models that are used, and the varieties of explainability that they offer.
\item Finally, we identified gaps and limitations and have suggested a roadmap for the future.

\end{itemize}

\section{Explainable AI in a Nutshell}
\label{bkg}
One of the common issues in the literature on Explainable AI is that there are several related concepts in this domain that are often used interchangeably in different articles. Therefore, in this section, we provide a brief description and the definitions of the terms that are widely used in Explainable AI jargon which help to understand the related literature and the concepts more precisely. This will also make the basis for the definition of XAI in SE, which will be used in this paper.

\subsection{Explainability and Interpretability}
The most common term which hinders the establishment of concepts in XAI is the interchangeable use of interpretability and explainability in the literature. Although these terms are closely related there are some works that identify and distinguish these terms. 

Though there are no clear mathematical definitions for interpretability and explainability, few works in this line attempt to clarify these terms. According to Miller~\cite{miller2019explanation}, model interpretability means ``the degree to which a human can understand the cause of a decision''. Another popular definition came from Doshi-Velez and Kim~\cite{doshi2017towards}, where they define it as ``the ability to explain or to present in understandable terms to a human''.
Based on these explanations \textit{Interpretability} is more like the cause and effect relationship of inputs and outputs of the model. 

In contrast, \textit{Explainability} is related to the internal logic and mechanics that are inside a machine learning algorithm. If the model is explainable, the human can understand the behavior of the model in the training and decision-making process. An interpretable model does not necessarily explain the internal logic of the ML model. This explains that interpretability does not exactly entail explainability, but an explainable model is always interpretable~\citep{gilpin2018explaining}. Therefore, it is necessary to address both interpretability and explainability aspects to better understand the ML model's logical behavior based on its inputs and outputs.

Regardless of this debate, since many studies have used these terms interchangeably and this SLR wants to cover all the related work, in both literature collection and analysis phases in this paper, we have ignored the differences between these two terms.

\subsection{Objectives of XAI}
\label{objectives}
Basically, explainability can be used to evaluate, justify, improve, or manage AI, or even learn from it. It is necessary to understand AI's risks and its failures by understanding the model's behavior.

In this section, we explain the main objectives of explainability in AI, in general (not limited to SE) so that later in the paper we can assess with aspects of generic XAI has been used and identified and useful in the SE domain. It's worth mentioning that these objectives are not separate concepts and may overlap in many aspects or be called with other names in different articles, but here, we gathered the most mentioned objectives in the literature.

An XAI method can aim to explain different aspects of an ML model. For instance, the practitioners may want to focus on the input data in an ML model to help them properly balance the training dataset. In another work, researchers may focus on the final output of the model to be able to provide a human-understandable explanation to the model's end user. In this section, we will go through some of these aspects of XAI and some suggested questions that can guide researchers' efforts, while focusing on these aspects.

\subsubsection{Justification}
One of the main objectives of explaining a model is to justify the model's decision in order to increase its credibility. This objective's  aim is to gain the trust of users or people who are affected by the AI model. For this purpose, it's not necessary to explain different components or algorithms of a model, but it's required to find the connection between inputs and outputs~\citep{gilpin2018explaining}. 

To answer this need, it's important to know why an instance gives a specific output. It would also be helpful to elaborate on the feature(s) of the instance that determines the prediction or in some cases, to explore the reasons for getting the same output for two different inputs. Sometimes it's important to know how much change and in what direction for an instance is required or tolerable, in order to see a different output. Being able to answer such questions will make it easier for the users to trust the models.

\subsubsection{Improvement}
Improvement is one of the most important and strongest inspirations behind many explainability efforts~\citep{xu2019explainable}. Working with black-box complex models, it is a common problem that the model starts to make wrong decisions but the reasoning behind it is unclear to developers or researchers. The problem can be due to imbalanced training data or an inadequate internal part or overfitting of the model on specific features. Providing an explanation can be helpful in these situations.

\subsubsection{Understanding}

Understanding a model is another motivation that is mentioned in the literature. It's a very general term but that usually pairs up with other objectives or makes them possible. Understanding a model may lead to its improvement, test its fairness, or increase the level of user confidence.

\subsubsection{Fairness}

\textit{Fairness} is a widely used term in the space of AI and ML. Hence, understanding the term fairness is important before discussing how fairness relates to Explainability of ML models. 
Broadly speaking, fairness is the quality or state of being fair or having impartial judgment.
Fairness has been discussed in many disciplines such as law, social science, quantitative fields, philosophy etc.~\citep{mulligan2019thing} 

% A few example definitions are discussed below:  

% \begin{itemize}
%     \item Law:  fairness includes protecting individuals and groups from discrimination or mistreatment with a focus on prohibiting behaviors, biases and basing decisions on certain protected factors or social group categories. 
%     \item Social Science: ``often considers fairness in light of social relationships, power dynamics, institutions and markets.''~\citep{mulligan2019thing} Members of certain groups (or identities) that tend to experience advantages. 
%     \item Quantitative fields (i.e. math, computer science, statistics, economics): questions of
% fairness are seen as mathematical problems. Fairness tends to match to some sort of criteria, such as equal or equitable allocation, representation, or error rates, for a particular task or problem. 
%     \item Philosophy:  ideas of fairness ``rest on a sense that what is fair is also what is morally
% right.''~\citep{mulligan2019thing} Political philosophy connects fairness to notions of justice and equity.

% \end{itemize}

In machine learning, fairness can be pursued, in order to prevent biases and discriminations that may happen in different parts of an ML. It can appear in the training data where the dataset is imbalanced, in algorithms where for instance using specific optimization functions may lead to biased decision-making, or in the output representation where wrong conclusions are drawn from an observation~\citep{mehrabi2021survey}.

XAI methods can significantly help researchers to consciously detect and eliminate biases and achieve fairness.

\subsubsection{Transparency}
\label{transparency}
Transparency is the opposite of black-boxness, that means when an ML model makes a decision, its whole process can be comprehended by a human~\citep{roscher2020explainable}. Full transparency is usually impossible to fulfill, but it can be achieved in three levels that are: simulatable, decomposable, and algorithmic transparency~\citep{love2022explainable}. 
Simulatable transparency means the model can be readily interpreted or simulated by a human, which means having a specific input, the user must be able to calculate the output having the model's parameters. Decomposable transparency is when smaller components of the model can be explained or explainable, separately. Finally, algorithmic transparent models are those whose training can get investigated mathematically. 

\subsection{Modality of Explanation}

Model interpretability can be achieved if the model is understandable without further explanation~\citep{zini2022explainability}.
According to the interpretability literature, the model understanding can be achieved by either considering inherently interpretable predictive models (i.e.; Linear regression, Random forest, Decision trees, etc.) or post-hoc interpretability approaches~\citep{danilevsky2020survey}.   

\subsubsection{Intrinsically Interpretable}

Intrinsically interpretable models are also called inherently interpretable models or self-explaining models. These models usually provide some level of transparency in their design and structure or they might generate additional outputs such as attention maps~\citep{mohammadkhani2022explainable}, disentangles representations, or textual or multi-model explanations alongside their predictions. Decision trees are one of the most famous self-explaining ML models. However, these inherently interpretable models often suffer from the accuracy-interpretability trade-off. 

\subsubsection{Post-hoc Interpretability}
This approach approximates the black-box model using a transparent surrogate model without changing the base model. In this approach, the explained model is treated like a black-box and can be obtained even without any access to its internal components which is ideal when the model is too complex. However, post-hoc methods can also be applied intrinsicly and be model-specific~\citep{carvalho2019machine}. Some of the most common post-hoc methods are LIME~\citep{ribeiro2016should} and SHAP~\citep{lundberg2017unified} which provide an explanation regarding defined features.

\subsection{Scope of Explanation: Local and Global Explainability}

Portability is an important aspect of post-hoc interpretability. It explains how far the explainer method is dependent on access to the internals of the global training model or the explained model. 

Explainability methods are called \textit{model-specific} or decompositional or white-box models, if the interpretability algorithm is restricted to explain only one specific family of models. In contrast, the algorithms which explain the output of different global models by considering the global model as a black-box, are called \textit{model-agnostic}~\citep{lipton2018mythos}.

According to the present deployments in the space of interpretability, post-hoc interpretability can be further classified as global interpretability and local interpretability based on the scale of interpretation.

\subsubsection{Global Interpretability}
In order to explain the model output globally in a comprehensive way, the knowledge about the trained algorithm, trained model, and input data would be sufficient. The basic idea behind the holistic global interpretability is understanding how the model makes decisions based on the entire feature set, parameters, and structures.  The outcomes of holistic-level interpretability are to help identify: (a) the features' importance levels, and (b) the feature interactions with each other.

\subsubsection{Local Interpretability} 
When discussing local interpretability for a single prediction, it considers a single instance and argues that how the model's prediction for the particular instance corresponds to that instance's features. The prediction will be on a linear or monotonic relationship of some features in the local context. For example, the condition of all patients might not linearly depend on the patients' blood pressure. But, if we look at one specific patient and examine that patient's condition, while watching the small changes of his  blood pressure, there is a higher chance of finding a linear relationship in the sub-region of our data. This is a reason to expect more accurate explanations from local explanations compared to global explanations.

\section{Methodology}

Our methodology for this review was mostly inspired by the guidelines proposed by Kitchenham for conducting an SLR~\citep{kitchenham2004procedures}. The proposed method includes three main stages: planning, conducting the review, and reporting the results.

The planning phase comprises two steps. Identifying the need for a review which is discussed in Section \ref{intro} and developing a review protocol in order to remove the researchers' bias in the reviewing process. Our review protocol's main components include research questions, search strategy, paper selection, inclusion and exclusion criteria, and data extraction. This review protocol is recurringly modified and evolved during the process. Finally, the third step is basically reporting the results and the analysis. 

The first two steps are discussed in the following section while the results are reported in Section \ref{results} and analyzed in Section \ref{discussion}.

\subsection{Research Questions}
Regarding the purpose of this SLR which is investigating the utilization of XAI methods for ML/DL models in software engineering, we formulated the following RQs:

\begin{enumerate}
    \item\textbf{RQ1: \rqone}
    \begin{enumerate}
        \item \rqoneOne
        \item \rqonetwo
        \item \rqonethree
    \end{enumerate}

    \item\textbf{RQ2: \rqtwo}
    \begin{enumerate}
        \item \rqtwoone
        \item \rqtwotwo
    \end{enumerate}

    \item\textbf{RQ3: \rqthree}
    \begin{enumerate}
        \item \rqthreeone
        \item \rqthreetwo
        % \item Which XAI techniques have been found useful?
    \end{enumerate}

\end{enumerate}

Each of these RQs are aimed to help us shed some light on different areas of our review. RQ1 is focusing on SE tasks that took advantage of any kind of XAI method to explain their models. While RQ2 is concentrated on the XAI methods and how they have been utilized. Finaly, RQ3 cares about how successful XAI has been in doing its specified task. Each of these RQs also has sub-questions that together, form the answer to their respective main question.

\subsection{Search Strategy}

The search strategy that is used in this work, starts by choosing the digital libraries to be searched. To perform the primary search phase, five main scientific publication databases in software engineering and AI have been selected, i.e., ACM Digital Library, IEEE, Springer, Wiley Online Library, and Elsevier.

% \begin{itemize}
% \item ACM Digital Library
% \item IEEE
% \item Springer
% \item Wiley Online Library
% \item Elsevier    
% \end{itemize}

We considered three main categories of ``Explainable AI'', ``Software Engineering'', and ``Artificial Intelligence'' to create our search strings. By integrating alternative terms and synonyms for each of them, we gathered a list of keywords in three categories as shown in Table~\ref{tab:search_terms}.

\begin{table*}[]
\centering
\renewcommand*{\arraystretch}{1.3}
\caption{Key search terms and their respective alternative search terms.}
\label{tab:search_terms}
\begin{tabular}{ll}
\hline
Key Search Terms          & Alternative Search Terms                                                                                                                                                                                                                                                                                                                                                                                                                        \\ \hline
Explainable AI            & \begin{tabular}[c]{@{}l@{}}Interpretable, Interpretability, Explainability, local model, global model, \\ model-agnostic, explainer model, explanations, interpretability algorithm, \\ black-box models, rule-based, counterfactual, causal\end{tabular}                                                                                                                                                                                       \\ \hline
Software   Engineering    & \begin{tabular}[c]{@{}l@{}}SQA, Software Analytics, defect prediction, security, vulnerability detection,\\ intrusion detection system, API recommendation, risk prediction, code \\ comprehension, code generation, code retrieval, code summarization, \\ bug-fixing processes, software testing, effort estimation, documentation \\ generation, clone detection, program synthesis, image to code, security, \\ program repair\end{tabular} \\ \hline
Artificial   Intelligence & \begin{tabular}[c]{@{}l@{}}Machine learning, classification, neural networks, deep learning, image\\ processing, text mining, text analysis, classification, clustering, rule \\ mining, association rules, NLP, Natural Language   Processing, \\ embedding, transformer, adversarial learning\end{tabular}                                                                                                                                    \\ \hline
\end{tabular}
\end{table*}

Using the boolean operators AND and OR, we formulated a search string that is:
(“Interpretable” OR “Interpretability” OR “Explainability” OR “local model” OR “global model” OR “model-agnostic” OR “explainer model” OR “explanations” OR “black-box models” OR “rule-based” OR “counterfactual” OR “causal”) 
AND
(“SQA” OR “Software Analytics” OR “defect prediction” OR “API recommendation” OR “risk prediction” OR “code comprehension” OR “code generation” OR “code retrieval” OR “code summarization” OR “bug-fixing processes” OR “software testing” OR “effort estimation” OR “documentation generation” OR “clone detection” OR “program synthesis” OR “image to code” OR “security” OR “program repair” OR “vulnerability detection” OR “intrusion detection system” OR “Malware detection”) 
AND
(“Machine learning” OR “classification” OR ”neural networks” OR “neural networks” OR “image processing” OR “text mining” OR “text analysis” OR “classification” OR “clustering” OR “rule mining” OR “association rules” OR “NLP” OR “Natural Language Processing” OR “embedding” OR “transformer” OR “adversarial learning”)

The OR operators will check if any of the search terms in one category exist in the paper and the AND operator will make sure that all three categories are found in the papers.

\subsection{Study Selection/ Inclusion and Exclusion Criteria}

\begin{figure}
\centering
  \includegraphics[width=0.9\linewidth]{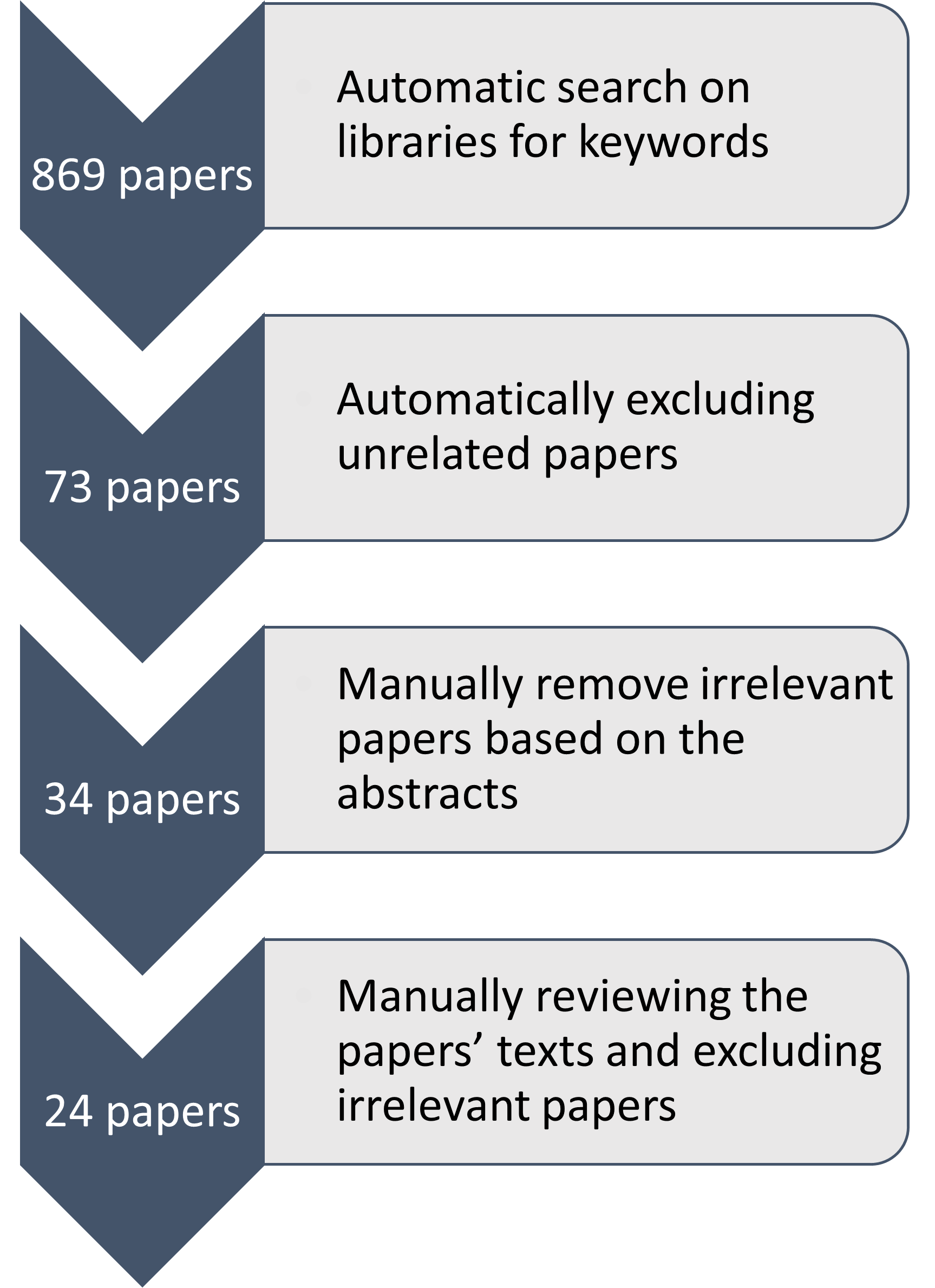}
  \caption{An overview of the searching strategy and its different stages with the number of studies in each step.}
  \label{fig:search_strategy}
\end{figure}

First, we started by searching a smaller version of the alternative search terms (more generic terms) on the full documents of papers in all of the specified databases with no time limit for the papers on the date ``06/textbackslash21/2022''. This ended up with hundreds of irrelevant papers as the result. So it was decided to limit the search to the title, abstract, and keywords of the papers. Given that the terms used in title, abstract, and keywords are more specific and not very generic (i.e., a generic word such as ``software engineering'' might not be used in the title, abstract, and keywords but the specific task e.g., ``defect prediction'' will be.), we augmented the search terms in the new search to the current list to make sure we do not miss any relevant paper. 
To make sure that our findings are not biased to a limited set of specific terms, we augmented the list iteratively (4 iterations), by manually checking the citations and references of the collected papers and updating the list to cover all relevant papers. 
% The current list was created after four rounds of snowballing~\citep{wohlin2014guidelines} by manually checking the citations and references of the collected papers and updating the list to cover all relevant papers. 
At this point, we could find a total of 869 papers which many of them were unrelated to SE and belonged to Health, Physiology, Medicine, or security. We used the search engine tools to filter the papers of unrelated fields that left us with 73 papers. 
% The last steady version (Table~\ref{tab:search_terms}) resulted in XX papers from each source are shown in Table~\ref{tab:final_list}.

Lots of irrelevant papers were still among the search results so in order to find and select the relevant papers, we manually reviewed the abstracts of all the 73 papers (this job was divided between four authors. Then one author validated all decisions) and included or excluded papers based on the following criteria:
\begin{itemize}
    \item Inclusion
    \begin{itemize}
        \item Full text papers published as journal or conference papers that comply with Interpretability and explainability definition provided in Section \ref{bkg}.
        \item Papers that are written in English Language.
        \item Papers that are related to software engineering.
        \item Papers that are available in full text.
        % \item Papers which explain interpretability and explainability of ML models based on tabular data and text data.
    \end{itemize}
    \item Exclusion
    \begin{itemize}
        \item Books, gray literature (theses, unpublished and incomplete work), posters, secondary or review studies (SLR or SMS), surveys, discussions and keynotes are excluded.
        \item Short papers where page count is less than 4 pages.
        % \item Papers with inadequate information to extract (Irrelevant Papers).
        \item Papers about Software Engineering but not discussing about interpretability and explainability.
        \item Papers about interpretability and explainability but not applying it on Software Engineering tasks.
    \end{itemize}
\end{itemize}

So if a paper meets all of the defined inclusion criteria and also does not meet any of the exclusion criteria, we included it in the final papers list. Finally according to all of these filters and criteria, we were able to extract 25 papers that totally matched our interests. One paper \footnote{https://ieeexplore.ieee.org/document/1374186} was discarded due to the problem of accessing its full version in the digital libraries. So there are 24 papers remained, which can be seen in Table~\ref{tab:final_list}.

\subsection{Data Extraction}

For the data extraction phase, we defined a checklist of items that could help us extract the required information from each paper to answer the RQs. The checklist includes both quantitative and qualitative properties. The 24 papers were divided between four authors and one author verified and consolidated the results. We defined 17 main properties that could help us answer the RQs as below:

\begin{enumerate}
    \item Publication details (Title, Authors, Published Date, Venue, Authors’ affiliation)
    \item What is the aim/ motivation/ goal of the research?
    \item What are the key research problems addressed?
    \item What phases of the SE are considered in the study?
    \item What SE tasks are under experiment in the study? (For papers that perform experiments and report results)
    % \item What are the data collection methods used in the study?
    % \item Does the research identify the most affected SE phase by human factors?
    % \item What are the existing domain models used in studies to identify human factors?
    \item Is the study conducted based on open-source data or data from industry?
    \item What are the ML/DL models and techniques considered in the study and how they have been evaluated?
    % \item Who are the target group that can benefit from the study?
    \item What are the XAI methods and techniques used in the study and how they are evaluated or represented (or visualized)?
    \item What is the scope and modality of the explanation offered in the study?
    \item What is the replication status of the experiments in the study?
    \item The number of participants used in the study where human evaluation was involved?
    \item What are the strengths and limitations of the study?
    \item What are the key research gaps/ future work identified by each study?
    % \item Does the research focus on identifying the relationship between different human factors?
    \item Is the explainability among the main focuses of the study or just a side-benefit
    % \item The number of citations of the study?
\end{enumerate}

% \subsection{Quality assessment}
\subsection{Data Synthesis}

Our data synthesis focused on summarizing and presenting the extracted data to convey meaningful information regarding the papers to address the RQs. In Section \ref{results}, we present all of our findings and results using tables, pie charts, bar charts, etc. in order to share insightful information from our analysis and studies.

\section{Results}
\label{results}
% This section summarizes the information \kla{info or results?} about the final selected papers after the search process and some\kla{some is an ambiguous term that must be removed throughout the paper} meta-analysis on their publication places and years.
the results of the final selected papers after the search process and the meta-analysis on their publication places and years.

\subsection{Selected Primary Studies} 

After filtering out or adding different research to our list of papers in multiple steps, finally we came up with a list of 24 papers that totally matched our search criteria. In Table\ref{tab:final_list}, the paper's title, name of authors, year and source of publication, and publication type (journal or conference) of all these final selected studies are summarized.

\subsection{Publication Statistics}

\begin{figure}
  \includegraphics[width=\linewidth]{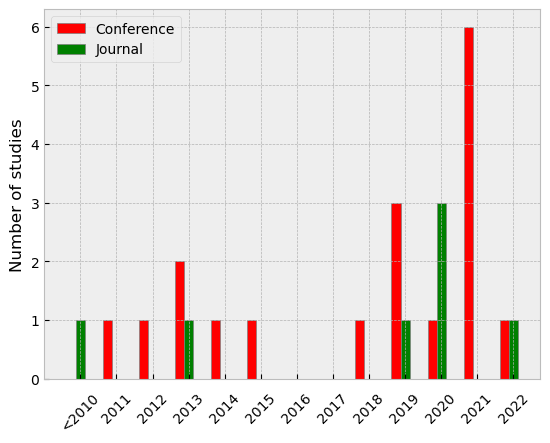}
  \caption{Distribution of selected studies over years and types of venues.}
  \label{fig:years_chart}
\end{figure}

\begin{table}[]
\centering
\renewcommand*{\arraystretch}{1.1}
\caption{Authors with more than one contribution in the selected studies.}
\label{tab:Authors}
\begin{tabular}{ll}

Author                       & \begin{tabular}[c]{@{}l@{}}\# of\\ papers\end{tabular} \\ \hline
Tantithamthavorn,   Chakkrit & 5                                                             \\
Jiarpakdee, Jirayus          & 3                                                             \\
Dam, Hoa Khanh               & 2                                                             \\
Pornprasit, Chanathip        & 2                                                             \\
Thongtanunam, Patanamon      & 2                                                             \\
Matsumoto, Kenichi           & 2                                                            
\end{tabular}
\end{table}
As shown in Figure\ref{fig:years_chart}, the first research in our selected studies goes back to 2008, published in Software Quality Journal. This study uses k nearest neighbor (kNN) and classification and regression trees generated based on the CART algorithm to perform the software effort prediction.  Even though this study does not directly mention the XAI concept, it touches upon the explainability aspect in an implicit manner by being able to find the dominant predictor variables and how they're statistically significant using Anova method. After that, there is a small number of XAI works in SE each year but then from 2019, we can see a trending pattern and popularity of related research.

% In terms of publication and venues, There are 4 studies published in Springer journals that half of them published in 2020 and the other two in 2008 and 2014. However, most of the publications from ACM and IEEE are published in conferences with only one and two journal papers for each of them respectively.
In the selected studies, a total of 75 authors contributed, while six mentioned in Table \ref{tab:Authors} are the authors in more than one paper. Also in terms of conferences and journals, as shown in Table \ref{tab:cons_jours}, only three journals and one conference had more than one paper in the selected studies, which confirms that this field is still a new domain in the SE community.

\begin{table}[]
\centering
\renewcommand*{\arraystretch}{1.1}
\caption{Conferences [C] and journals [J] and number of papers published in each venue. The full name of each venue can be found in Table \ref{tab:final_list}}
\label{tab:cons_jours}
\begin{tabular}{ll}
Conference/Journal                                      & \begin{tabular}[c]{@{}l@{}}\# of\\ papers\end{tabular} \\ \hline
RAISE [C]      & 2                                                      \\
MSR [C]                                    & 2                                                      \\
Transactions on Software   Engineering [J] & 2                                                      \\
Software Quality Journal [J]               & 2                                                      \\
Empirical Software Engineering   [J]       & 2                                                      \\
Symposium on Cloud Computing   [C]         & 1                                                      \\
TOSEM [J]                                  & 1                                                      \\
IUI [C]                                    & 1                                                      \\
ICSE-NIER [C]                                   & 1                                                      \\
Symposium on Applied Computing   [C]       & 1                                                      \\
ICSE-Companion [C]                         & 1                                                      \\
ESEC/FSE [C]                               & 1                                                      \\
RE [C]                                     & 1                                                      \\
ICMLA [C]                                  & 1                                                      \\
ASE [C]                                    & 1                                                      \\
ICECS [C]                                  & 1                                                      \\
ISSRE [C]                                  & 1                                                      \\
DSA [C]                                    & 1                                                      \\
ICTAI [C]                                  & 1                                                     
\end{tabular}
\end{table}

% Also before 2018, there was only one paper in ACM, but in 2019 alone, there were 3 publications showing the increasing interest of XAI in the venue. However, IEEE has shown much greater interest in the topic as in 2021, they had 5 publications related to XAI. \kla{what is the key message of this paragraph?}

Another interesting observation is that before 2018, most of the publications were in AI conferences or journals that weren't specific to SE, while after that, all of the publications belong to SE-specific venues which shows the rising interest of the SE community in the field.

% \begin{figure}
%   \includegraphics[width=\linewidth]{figs/venue.png}
%   \caption{Distribution of selected studies in different venues.}
%   \label{fig:venues}
% \end{figure}

\section{Synthesis and Analysis}

\label{discussion}

\subsection{RQ1 Results: \rqone}

In our selected studies, we surveyed different SE activities and SE tasks that XAI methods and studies have been focused on or utilized. We organized the tasks and activities based on the Software Development Life Cycle (SDLC) stages~\citep{yang2022survey} and in this section, we are going to present and analyze the results to answer RQ1 of this study.

\subsubsection{RQ1.1 results: \rqoneOne}
\label{missing_areas}

\begin{figure}
\centering
  \includegraphics[width=0.5\textwidth]{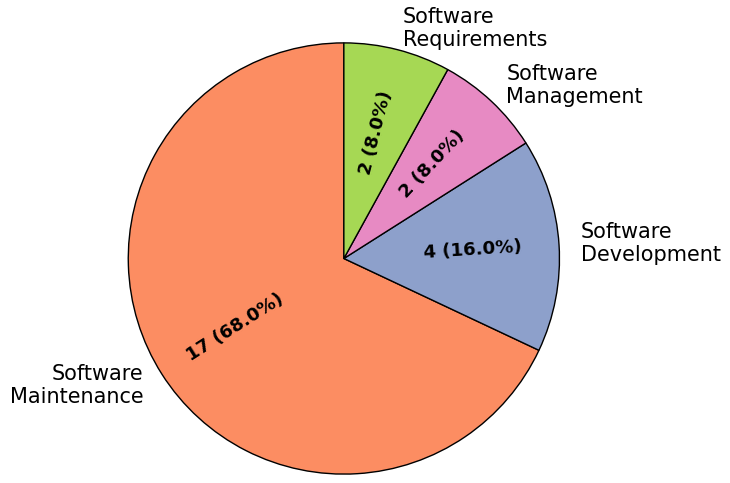}
  \caption{Distribution of studies per SDLC stage in the selected studies.  Reported by \#Papers (\% proportions over all).}
  \label{fig:activities}
\end{figure}

The most attractive activity for the researchers was software maintenance with 68\% of the tasks being used for interpretation as shown in Figure \ref{fig:activities}. After that, software development has the largest share with 16\%, and software management and software requirements each with two tasks in all of the studies have 8\%. It's noteworthy that among the six stages of SDLC, software design and testing are two activities that have no presence in all of our selected studies.

% Please add the following required packages to your document preamble:
% \usepackage{multirow}
\begin{table}[]
\renewcommand*{\arraystretch}{1.3}
\caption{The break-down of SDLC stages to tasks and their number of appearance in the selected studies.}
\label{tab:Tasks}
\begin{tabular}{cll}
\hline
\multicolumn{1}{c}{Activity}   & Task                                                                             & \# \\ \hline
\multirow{2}{*}{Requirement}   & Requirements engineering                                                         & 1  \\ \cline{2-3} 
                               & Requirements quality                                                             & 1  \\ \hline
\multirow{4}{*}{Development}   & Code translation                                                                 & 1  \\ \cline{2-3} 
                               & Code autocompletion                                                              & 1  \\ \cline{2-3} 
                               & Natural language to code                                                         & 1  \\ \cline{2-3} 
                               & Code Summarization                                                               & 1  \\ \hline
\multirow{6}{*}{QA \& Maintanance}   & Quality assessment                                                               & 1  \\ \cline{2-3} 
                               & Defect prediction                                                                & 11 \\ \cline{2-3} 
                               & Code reviews                                                                     & 1  \\ \cline{2-3} 
                               & \begin{tabular}[c]{@{}l@{}} Reliability prediction\end{tabular}       & 1  \\ \cline{2-3} 
                               & \begin{tabular}[c]{@{}l@{}}Technical debt detection\end{tabular} & 1  \\ \cline{2-3} 
                               &
                               \begin{tabular}[c]{@{}l@{}}Valid bug report detection\end{tabular} & 1  \\ \cline{2-3} 
                               &
                               Variable misuse detection                                                        & 1  \\ \hline
\multicolumn{1}{c}{Management} & Effort Prediction                                                       & 2  \\ \hline
\end{tabular}
\end{table}

A particularly noteworthy finding is the high ratio of the defect prediction task compared to all other tasks in the studies. This task alone has the lead with 44\% of the total number with a meaningful gap from the next task, which is effort prediction with only two studies and the other tasks each one being utilized only once. Also in general, QA and software maintenance tasks have the highest diversity with having seven different tasks. In Table\ref{tab:Tasks}, we can see different tasks in each activity domain's broken-down.
 
Defect prediction is basically a classification problem to tag a code snippet, class, etc. as faulty or not faulty. This makes it a very ideal task for machine learning algorithms and also for XAI. It's also a very important task that ML models have shown promising results in.

We also examined a survey~\citep{yang2022survey} on deep learning models for software engineering tasks and interestingly, we found 27 different tasks that already have taken advantage of black-box DL models, and yet, no explanation method has been used on them. Tasks such as ``Code generation'', ``Code comment generation'', ``Test case generation'', ``Bug classification'', etc. This can show the big gap between ML applications in SE and the utilization of XAI in the field.

Furthermore, we observed that among the reviewed studies that include an empirical evaluation, 16 papers only used open-source data, three papers only used publicly unavailable industrial data~\citep{16,31,7}, while two papers used both~\citep{9,12}. 
The data type in all the studies are either textual which can be source code or NL, or tabular which is a set of extracted features.

\subsubsection{RQ1.2 Results: \rqonetwo}

\begin{figure*}
\centering
  \includegraphics[width=\textwidth]{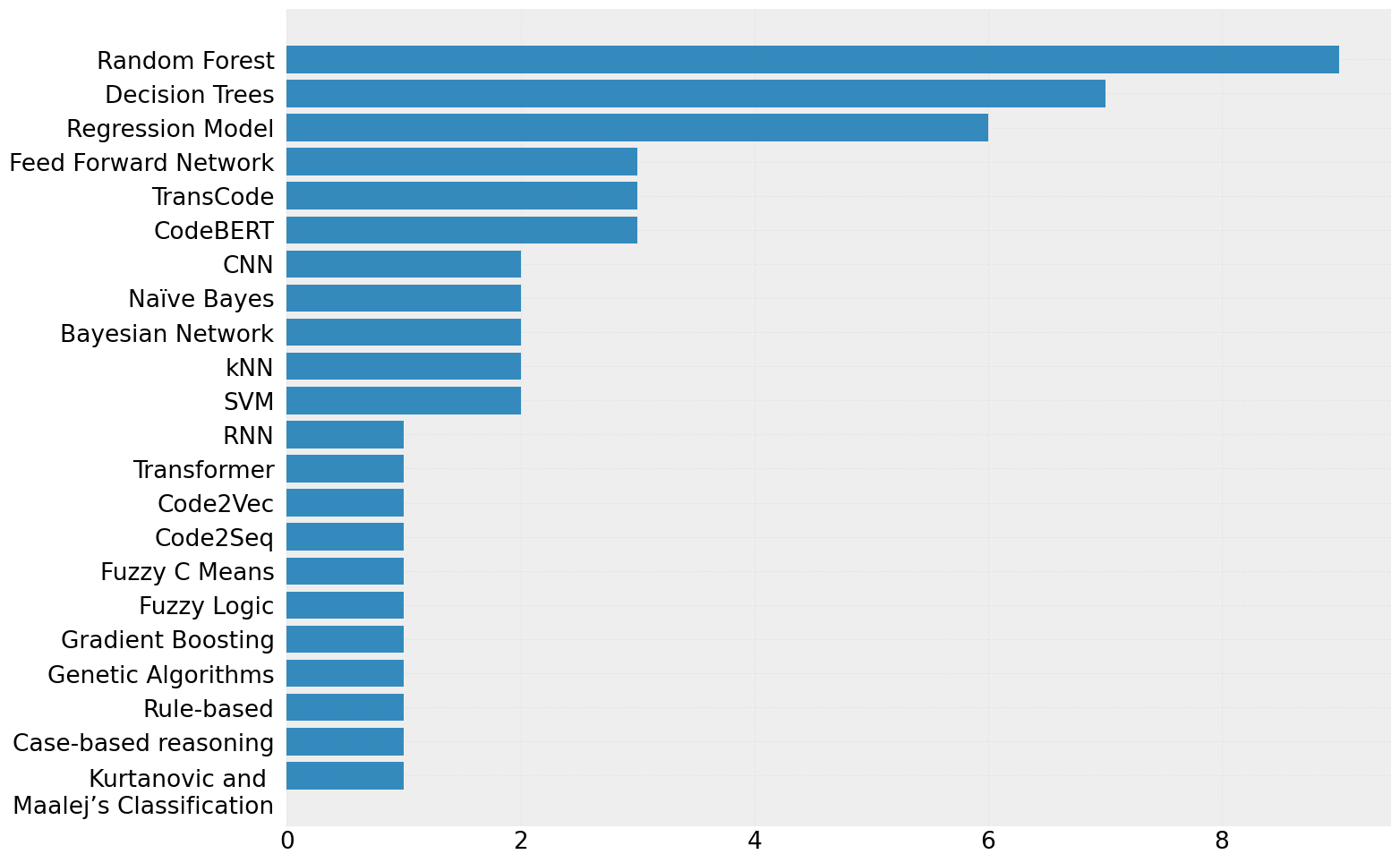}
  \caption{Number of experiments each ML model have been used in the selected studies}
  \label{fig:ml_algorithms}
\end{figure*}

To answer this question, we analyzed different ML models and the number of times that they have been used in the selected studies. As the sorted results show in Figure \ref{fig:ml_algorithms}, while many models are rare and have only one use case, models like random forest and decision tree have been very popular with nine and seven uses. This is probably due to the fact that these classic models are usually being used as the baselines in many papers and also the fact that these models offer a level of self-explainability so it is expected to see them more in our selected studies.

The next popular models are regression models that are used for both regression problems (such as effort prediction) and classification problems (such as defect prediction). Regression models are used six times in total, while four of them are logistic regression, and linear and robust regression each have only one use-case.

Neural networks are also favored by having a total of 15 times being experimented, with adequate diversity among themselves. More complex code models like Code2Vec, Code2Seq, TransCode, and CodeBERT can be found in the studies, as well as simpler networks like feed-forward neural networks, CNN, RNN, or basic transformer models are also among the explained models.

In Table \ref{tab:ml-model} ML models are shown alongside the tasks they were used to solve. Models like the random forest or regression models are mostly utilized for defect prediction. The popularity of random forest for defect prediction is probably due to the fact that it's a self-explaining model which is a good fit for this specific task with high accuracy. Also, it is interesting to see the diversity of tasks that take advantage of the decision tree as a simple yet effective model.

DNN code models like Code2Vec, Code2Seq, Transcode, and CodeBert are always used for generation-based tasks such as code summarization or code translation which is understandable due to the complexity of these tasks and traditional models' incompatibility with them.

Regarding a large number of defect prediction studies, we also focused on those papers more specifically. Among the 11 works focused on this task, there is quite a diversity of ML models being used and even some studies cover different models for the purpose of their research. As mentioned, random forest is the most popular method that has been used in eight different papers either as the main model or as a benchmark to compare with. Logistic regression and deep learning neural networks are the second most popular models being used in five and three papers. There are other models that only have been used in two or one study. Models such as Support Vector Machine (SVM), k-Nearest Neighbours, and C5.0 Boosting. In these defect prediction tasks, standard evaluation metrics for ML models such as precision, recall, F1 score, and AUC are commonly used.

\begin{figure}
\centering
  \includegraphics[width=0.5\textwidth]{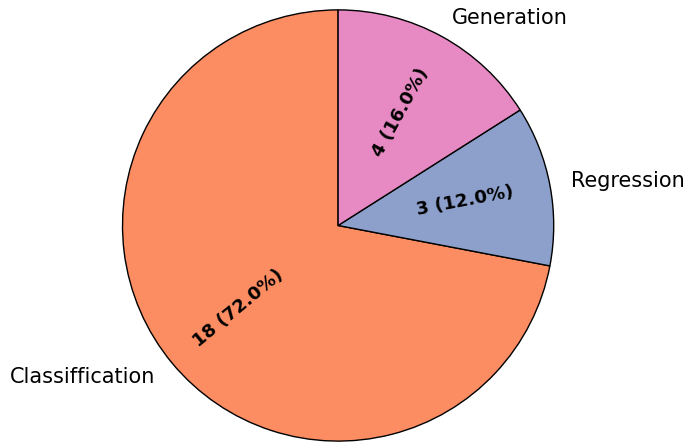}
  \caption{Distribution of SE task types in selected studies. Reported by \#Papers (\% proportions over all).}
  \label{fig:algorithms}
\end{figure}

We also analyzed the SE tasks that these ML models have been applied to in terms of their ML problem type. As shown in Figure\ref{fig:algorithms}, 18 tasks (more than 65\% of the tasks experimented in the selected studies) fall into the classification problems. While only three tasks are considered regression problems (reliability prediction, requirements quality,  and effort prediction), and four can be called generation problems (code translation, code autocompletion, natural language to code, and code summarization).

Classification tasks are among the simplest models to explain because of the fixed number of possible outputs that leads to a more straightforward evaluation, thus it makes sense for the researchers to focus on them. On the other hand, generation tasks are among the hardest since the output is usually harder to evaluate objectively. This gets more interesting when we see that three of those generation tasks are actually from one paper~\citep{5} that is doing a human evaluation of XAI methods for SE tasks. We will discuss the challenges of evaluation metrics in Section \ref{evaluation_metric}.

\begin{table*}[]
\centering
\caption{Different ML models and the tasks they have been used for and number of uses}
\label{tab:ml-model}
\renewcommand*{\arraystretch}{1.4}
\begin{tabular}{lll}
\hline
ML model                                                                          & Task (\# of Uses)                                                                                                                                                     & \begin{tabular}[c]{@{}l@{}}Total \\ \# of uses\end{tabular} \\ \hline
Random Forest                                                                     & Defect prediction (8) , Code reviews                                                                                                                                  & 9                                                           \\ \hline
Decision Tree                                                                    & \begin{tabular}[c]{@{}l@{}}Effort prediction, defect prediction (2), Quality assessment, \\ Requirements quality, Requirements engineering, Code reviews\end{tabular} & 7                                                           \\ \hline
Regression Model                                                                  & Defect prediction (5), Requirements engineering                                                                                                                       & 6                                                           \\ \hline
TransCode                                                                         & Code translation, Code autocompletion, Natural language to code                                                                                                       & 3                                                           \\ \hline
CodeBERT                                                                          & Code translation, Code autocompletion, Natural language to code                                                                                                       & 3                                                           \\ \hline
Feed Forward Network                                                                    & Defect Prediction (2), Quality assessment                                                                                                                             & 3                                                           \\ \hline
Naïve Bayes                                                                       & Defect prediction (2)                                                                                                                                                  & 2                                                           \\ \hline
kNN                                                                               & Defect Prediction, Requirements engineering                                                                                                                           & 2                                                           \\ \hline
SVM                                                                               & Quality assessment, Defect Prediction                                                                                                                                 & 2                                                           \\ \hline

CNN                                                                               & self-admitted technical debt detection, Valid Bug Report                                                                                                              & 2                                                           \\ \hline
Bayesian Network                                                                  & Quality assessment, Defect Prediction                                                                                                                                 & 2                                                           \\ \hline
RNN                                                                               & Variable Misuse detection                                                                                                                                             & 1                                                           \\ \hline
Fuzzy Logic                                                                       & Software Reliability Prediction                                                                                                                                       & 1                                                           \\ \hline
Transformer                                                                       & Variable Misuse detection, Defect prediction                                                                                                                          & 1                                                           \\ \hline
Code2Vec                                                                          & Code Summarization                                                                                                                                                    & 1                                                           \\ \hline
Code2Seq                                                                          & Code Summarization                                                                                                                                                    & 1                                                           \\ \hline

Fuzzy C Means                                                                     & Defect Prediction                                                                                                                                                     & 1                                                           \\ \hline
\begin{tabular}[c]{@{}l@{}}Kurtanovic and Maalej’s \\ Classification\end{tabular} & Requirements Classification                                                                                                                                           & 1                                                           \\ \hline
Gradient Boosting                                                                 & Defect prediction                                                                                                                                                     & 1                                                           \\ \hline
Genetic Algorithms                                                                & Software Reliability Prediction                                                                                                                                       & 1                                                           \\ \hline
Rule-based                                                                        & Quality assessment                                                                                                                                                    & 1                                                           \\ \hline
Case-based Reasoning                                                              & Quality assessment                                                                                                                                                    & 1                                                           \\ \hline

\end{tabular}
\end{table*}

\subsubsection{RQ1.3 Results: \rqonethree}
\label{objectives_of_researches}
In Section \ref{objectives}, we demonstrated different objectives of XAI. In our analysis of the selected studies of this work, we were interested to extract different proclaimed objectives of different studies. We found three main objectives explicitly or implicitly mentioned in the selected studies as: accountability, fairness, credibility, understanding, transparency, and improvement. Accountability and credibility have been used interchangeably in different works, meaning the reliability of the model for its users. Note that these objectives are not mutually exclusive and some papers follow multiple goals so the numbers add up to greater than the number of papers.

\begin{figure}
  \includegraphics[width=\linewidth]{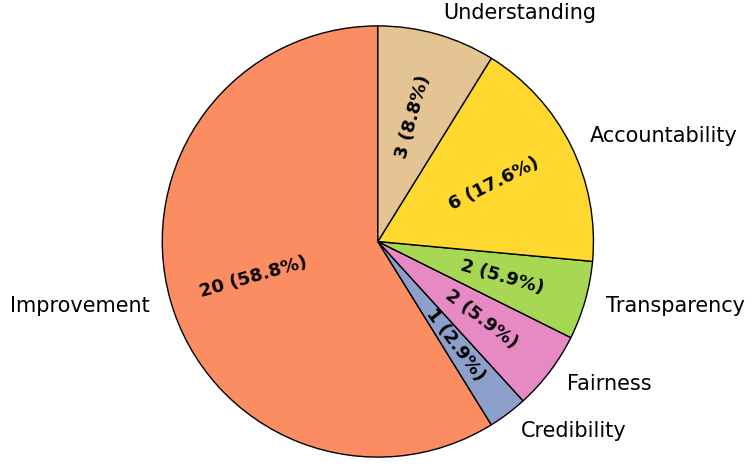}
  \caption{The XAI objectives stated (explicitly or implicitly) in the selected studies along with the number of papers that have mentioned them. Reported by \#Papers (\% proportions over all).}
  \label{fig:goals}
\end{figure}

As shown in Figure \ref{fig:goals}, 20 papers that is the vast majority of studies have claimed improvement as (at least) one of their purposes, accountability in six, and understanding in three different papers have been mentioned. transparency and fairness are less common objectives while being the aim of two papers, and credibility has been mentioned only once.

Accountability and fairness are usually used in reviews/surveys~\citep{3, 5, 6} together, but the other objectives are coupled with the improvement in different research.

It's worth mentioning that while most of the objectives have been said with the same meaning as what is discussed in Section \ref{objectives}, improvement is an ambiguous term, used for multiple intentions. In the selected studies some researchers consider the explanation as it is, as the improvement~\citep{9}. In the other words, they believe that the fact that their model offers an explanation is an improvement compared to other models. Tree-based models that extract rules for their decision-making and present them are among this category. 

Meanwhile, some studies have provided model-agnostic XAI tools and methods to help other researchers to understand and improve their models in the future~\citep{15} or even have used XAI methods to literally improve the results of models, especially in defect prediction. We will discuss them later in Section \ref{XAI_objectives} where we discuss the helpfulness of XAI for SE models.
\finding{\textbf{Answer to RQ1:}
% Our findings show a considerable difference among the explored tasks by the researchers as almost half of the experiments in the selected studies were conducted on the defect prediction task, while other tasks have only been experimented once or twice. There are many untouched SE tasks that have leveraged ML models, while have not benefitted from XAI.
Among different stages of SDLC, QA and maintenance have been the most popular among the XAI researchers, and this popularity is hugely focused on the defect prediction task. This is likely due to 1) the popularity of this task among researchers and also 2) the adequacy of self-explaining traditional models like decision trees for these tasks.

Our analysis also shows that, due to their inherent interpretability, classic models, like random forest, decision tree, regression models, and naive Bayes models, are among the most common ML models in the selected studies. DNN models are also another center of attention in terms of ML models, however generation-based tasks that are one of the main strengths of these models are still unexplored.

}

\subsection{RQ2 Results: \rqtwo}

In this section, we focus on the XAI methods that have been used to explain the SE models that we discussed in previous sections. This will include post-hoc XAI methods and self-explaining models.

\subsubsection{RQ2.1 Results: \rqtwoone}

As we discussed in Section \ref{transparency}, there are two types of XAI approaches: (a) models that offer some level of explanation alongside the output, and (b) post-hoc XAI techniques that are applied on the model afterwards. In this section, we are discuss both cases among our selected studies. According to our analysis, 15 self-explaining models were used in the studies that need no or a small amount of effort to interpret while only 9 models were using post-hoc methods.

Tree-based models are the most popular self-explaining models among these studies with using a diverse set of algorithms. Decision trees using the C4.5 algorithm or its developed version, the PART algorithm that uses fuzzy logic to handle the classification task~\citep{7, 13, 27, 31, 32} or random forests are widely used in our primary selected studies~\citep{16, 18, 19, 25,28}.

The explanation offered in tree-based models is usually a set of rules or highlighted features that are quite self-explanatory for the user and usually need no further processing. Thus, the explanation in these models are either in form of feature summary statistics, or feature summary visualization using plots and graphs.

\label{internal}
Deep neural networks are another observed type of ML models that even though are known as black-box models, but some researchers believe they have internal features that can be used for interpretation, after proper processing. For instance,~\citep{4} has used backtracking of a trained CNN model to extract the key phrases of the input and discover meaningful patterns in the code comments. As an explanation, they were able to find some 1-gram, 2-gram, and 3-gram patterns in natural language format that are compared to human-extracted patterns. Their model is able to cover most of the benchmark and also offers further comprehensive patterns with more diversity of vocabulary. In another study~\citep{23}, the authors also use backtracking of a CNN model but for the valid bug report determination. As the input of this task is in NL format, it is very similar to classification tasks in NLP and it generates a natural language explanation.

Transformer models are also among the promising DNN models that have outperformed state-of-the-art models in some SE tasks by leveraging the attention mechanism. There is a controversial debate about the validation of transformer model's attention mechanism as an explanation method, in recent years. Among the selected studies, there is one that have used the self-attention mechanism of a transformer as an explanation~\citep{1}. They claim to find the importance of the input tokens by constructing the attention matrices. The generated matrix as well as a respective saliency map is the explanation that they present, but no evaluation is discussed.

%There is also a very recent work on this area where the authors have explored the possibility of using the attention mechanism in transformer models, but instead of using the self-attention, they have utilized the end-to-end decoder-encoder attention scores. They showed the importance of those scores in the decision-making process of the pre-trained code models and then used them to offer explanations for generation-based tasks. However, since this work is not published yet (only on Arxiv), we didn't include it in our selected studies.~\citep{mohammadkhani2022explainable}

Looking at the post-hoc XAI methods, most techniques that are used in SE are adopted from the NLP domain. LIME (Local Interpretable Model-Agnostic)~\citep{ribeiro2016should} is one of the widely used XAI methods in NLP that also showed very useful in SE. In the reviewed papers, four different articles have used LIME directly in their studies~\citep{18,19, 25,28}. The method works based on feeding a black-box model with perturbed input data and observing the changes that happen on the output weights. This straightforward mechanism makes it ideal for NL and PL data. Perturbation can be defined on multiple levels from files in a commit to tokens in a code snippet and the results will be scores for the defined features (lines, tokens, etc.).

As we mentioned, LIME works based on input perturbation by generating synthetic instances($n$) similar to a test instance ($x$) by a random perturbation method. The level and granularity of this perturbation depend on the task and objectives. In two of the studies that use LIME~\citep{18, 28}, code tokens are considered as features that are supposed to be perturbed, while another study~\citep{25} uses tabular features of a code commit made by the developers (e.g. number of files modified, number of added or deleted lines, Number of developers who have modified the files, etc.).  After the generation of synthetic instances, a defect prediction model is used to label them, and then, based on the predictions, a score between -1 to 1 is assigned to each feature. A positive score means a positive impact on the prediction and vice versa. 
 
This model-agnostic and easy mechanism has led to the popularity of LIME among users, as another study, which is focused on the practitioners’ perceptions of XAI for defect prediction models~\citep{19}, claims that LIME gets the highest appeal among different XAI methods in terms of information usefulness, information, insightfulness, and information quality with more than 66\% agreement rate. 

The second favorite method is ANOVA (ANalysis Of VAriance), which is a statistical method that can be used to measure the importance of factors in a model. It does this by calculating the improvement in the Residual Sum of Squares (RSS) when a feature is added to the model. In our selected primary studies, This model-agnostic technique is used three times~\citep{19, 21, 33}. In addition, LIME and ANOVA have been selected as the first and second most preferred XAI methods, respectively, in a human evaluation of 50 software practitioners.

SIVAND is another model-agnostic XAI method that works based on Delta Debugging and reducing the input size, without changing the output~\citep{10}. The big picture is very similar to LIME but instead of just finding the most important atomic unit (token or character), it removes redundant units so the input gets smaller in size, but the prediction remains the same. They have tested their method on Code2Vec, Code2Seq, RNN, and Transformer models on two tasks of Method Name Prediction and Variable Misuse Detection and have a qualitative example-based evaluation of their models. For the Transformer model, they validate their performance by comparing the similarity of SIVAND and attention scores. 

PyExplainer is another XAI method inspired by LIME and focused on Just In Time (JIT) defect prediction. It is a model-agnostic local rule-based technique that challenges the LIME's random perturbation mechanism by generating better synthetic neighbors, and explanation more unique and consistent. According to the mentioned criteria, PyExplainer outperforms LIME on Random Forest and Logistic Regression models for defect prediction.

\begin{figure*}
\centering
  \includegraphics[width=0.8\textwidth]{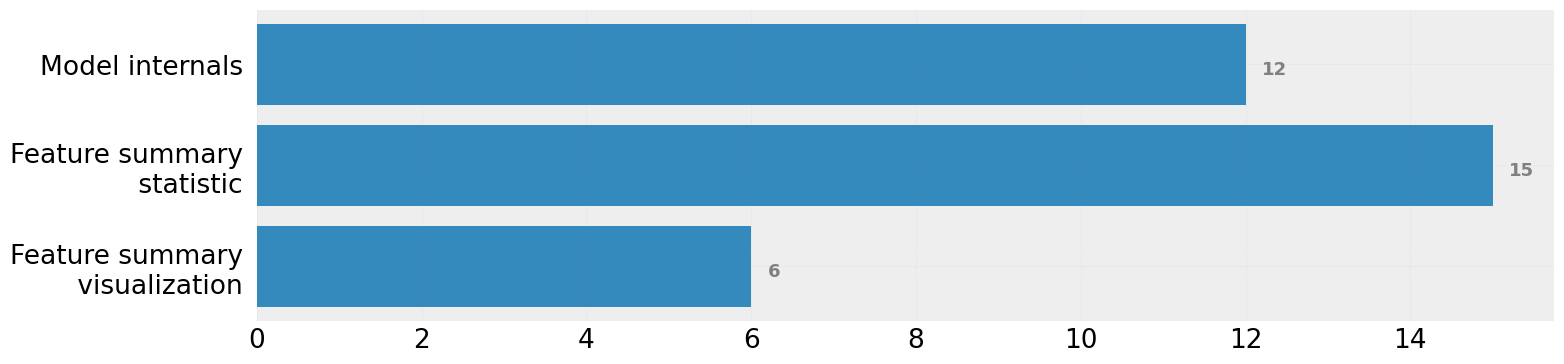}
  \caption{Different types of explanation offered in the selected studies.}
  \label{fig:type_of_explanation}
\end{figure*}

Beside these model-agnostic methods, there are some other XAI techniques that are used for interpreting specific ML models. For instance, deconvolution is a method that exploits the structure of CNNs to extract the importance of features\cite{4}. Flexible Code Counter/Classifier(CCFlex) is also an interesting model that was originally designed to classify lines of code but is used for code review and finding guideline violations\cite{34}.

Some other XAI methods are also mentioned in a study asking about software practitioners’ perceptions of the goals of XAI methods for defect prediction models\cite{19}. As mentioned earlier, LIME and ANOVA were the most preferred methods, while other techniques like Variable Importance, Partial Dependence, Decision Tree, BreakDown, SHAP,
and Anchor were less favorable.
% \subsubsection{domain-specific XAI techniques for SE}
\subsubsection{RQ2.2 Results: \rqtwotwo}
We analyzed the selected studies in terms of the type of explanation that they offer. Following the previous discussion about self-explaining models, many models deliver ``Model internals'' as the explanation. Extracted rules and structures of tree-based models or the attention matrix of Transformer models are in this category. However, some works go further and more than the local explanations, presenting the ``Feature summary statistic'' or ``Feature summary visualization'' of their explanations with different visualization methods. The distribution of each type is shown in Figure \ref{fig:type_of_explanation}. As it can be understood from the plot, some models have one, while others offer more types of explanation.

\begin{figure*}
\centering
  \includegraphics[width=0.8\textwidth]{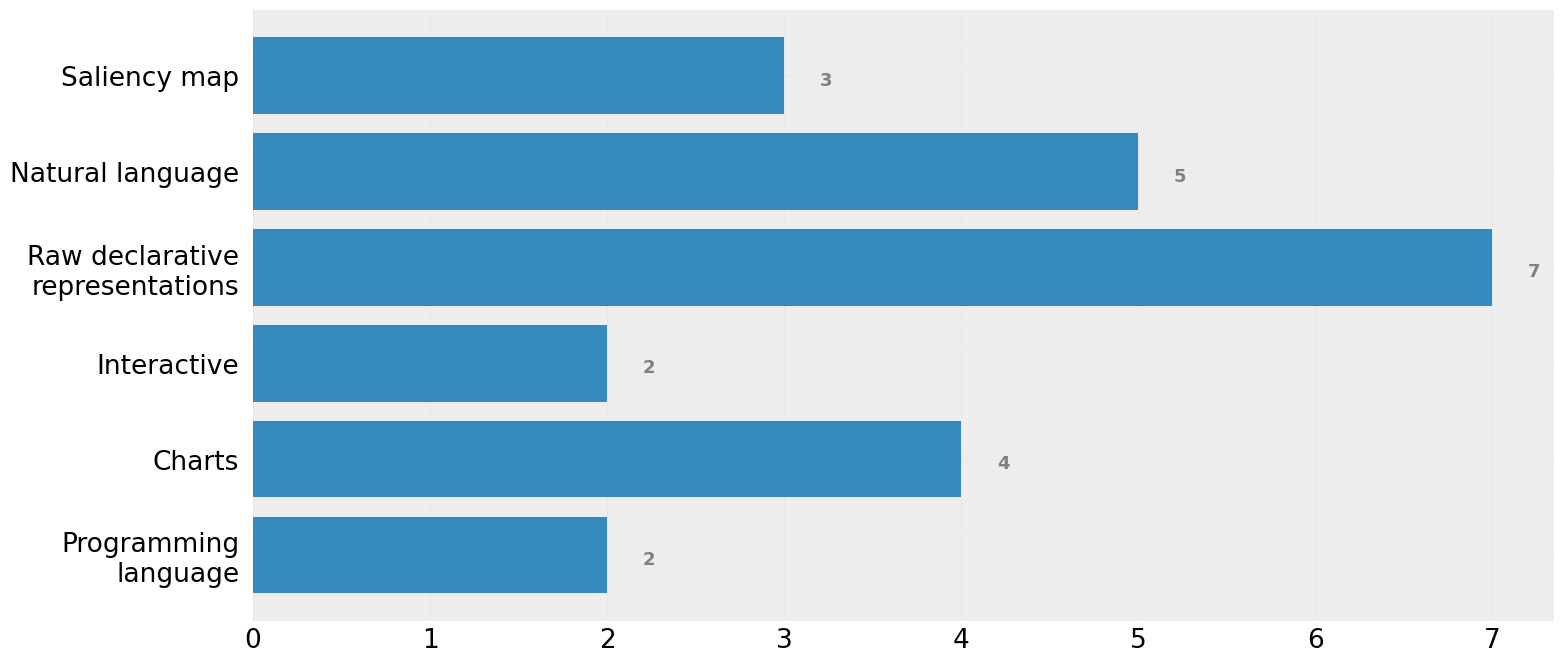}
  \caption{Distribution of different visualizations used as explanation methods in the selected studies.}
  \label{fig:type_of_visualization}
\end{figure*}

In the category of ``Feature summary visualization'', there are multiple XAI methods presented in the studies. While there are works that are satisfied by ``Raw declarative representations'' of their interpretations, many XAI methods implement more informative ways. ``Natural language explanation'' is used five times, while ``Saliency maps'' and ``Plots and charts'' each are used three times. There are also two methods that present their results in interactive tools to the users. The distribution of different visualization techniques in the studies is illustrated in Figure\ref{fig:type_of_visualization}.

\finding{\textbf{Answer to RQ2:}
Our investigation shows that beside different ML models that have a level of interpretebility and are quite common for SE tasks (such as random forest and decision trees), there is a variety of model-agnostic XAI methods that are used in SE to explain black-box models. LIME which is a perturbation XAI method is the most favorite method in the selected studies, while ANOVA is the second most-used method. There are also some other proposed methods used in the studies such as deconvolutioning for CNN models, or SIVAND which has similarities with LIME.  

The explanations offered by XAI methods can get categorized as ``Feature summary statistic'', ``Model internals'', or ``Feature summary visualization'' that have been used in the studies, respectively in descending order.
}

\subsection{RQ3 Results: \rqthree}
In this section, we are going to focus on the helpfulness of XAI methods that are used for SE tasks.
% which works are focused on XAI or they have a model and offer the explainability as well

\subsubsection{RQ3.1 Results: \rqthreeone}
\label{XAI_objectives}

In Section \ref{objectives_of_researches}, we discussed different objectives of research in our selected studies. Some of them have used explanation as an additional output for their users, while a few others used the XAI methods to improve the performance of the models.

One interesting representation of these improvements can be seen in defect prediction. While many defect prediction models only classify if a file or a class is faulty, some researchers have taken advantage of XAI methods to specifically find the buggy line or tokens of the code~\citep{18, 19, 28}. This is interesting since it improves defect prediction's practicality and its potential for adoption by industry.
%converts the defect prediction problem to another SE task, fault localization which is considered a software testing activity, unlike defect prediction.

SIVAND is another successful XAI method, where its finding improves the performance of the code models it is applied to, and also offers valuable knowledge about them~\citep{10}. They are able to considerably reduce the size of the inputs, and yet achieve the same accuracy. This means a great reduction in the models' required time to make a prediction. Furthermore, they offer interesting knowledge about code models by claiming the presence of alleged patterns in them. For instance, they believe while Transformer models can understand more ``global'' information in codes, RNN models are prone to local information. They also believe these models are quite vulnerable to perturbations as they rely on too few features.

Another noteworthy and exemplary use of XAI can be seen in another work~\citep{23} where by backtracking a trained CNN network and manually inspecting the explanation, the authors are able to find valid bug report patterns. Further analysis of the identified bug reports, they verify the importance and effect of XAI on their model and the problem.

\subsubsection{RQ3.2 Results: \rqthreetwo}
\label{evaluation_metric}
Looking at the evaluation of the XAI methods (not the ML models) across the selected papers, we can see that the struggle is apparent. While studying these selected papers, we observed that there is little consistency among selected papers in terms of evaluation. Many studies have totally ignored the evaluation phase and only offer some visualization for their XAI with no evidence of its quality or reliability. 

Most existing XAI evaluation methods~\citep{14},~\citep{4},~\citep{34} are qualitative and use human subjects to assess the explanations. 
The problem with this approach, specially with small scale subject pools, is that they are heavily biased and may not be generalizable to other users.
%In the lack of objective evaluation, some papers qualitatively analyze their results~\citep{14} while some others use human evaluation~\citep{4},~\citep{34}. Even 
In an attempt to find out how thoroughly evaluate XAI in SE, in one work~\citep{5} the authors asked 43 software engineers in 9 workshops to express their explainability needs for some specific tasks.

With respect to qualitative assessment of XAI methods in SE, there are some works that based on their task (e.g., defect predicion), are able to define more common metrics such as Mean Squared Root (MSR), or Wald test~\citep{21}.
In a few other studies, researchers defined innovative metrics or used some less-known metrics from generic the XAI domain. For instance, in one work~\citep{15} the authors measure the percentage of unique explanations generated by XAI method as a metric. Another work~\citep{13} utilizes the average length of the rules that their XAI extracts and another study~\citep{25} uses Variable Stability Index (VSI) and Coefficient Stability Index (CSI), as metrics. 

\finding{\textbf{Answer to RQ3:}
Our findings show that some XAI methods were quite helpful and were applied either for providing proper explanations to researchers that led to finding interesting patterns in the model's decision-making, or for increasing the granularity of models' results, leading to the higher precision for the respective SE task.

We also noticed some confusion and a lack of standard routines among the researchers in terms of evaluating XAI methods that are usually compensated by defining innovative and task-specific metrics or using qualitative evaluation by a human. 
}

% \subsubsection{XAI techniques proved useful}

\section{Challenges and road ahead}

%This study aimed to systematically analyze the literature on XAI for software engineering. We have searched and selected those researches that fulfilled our criteria. We analyzed them from different aspects and tried to shed some light on what have been done so far regarding the problem at hand. 
In general, there are valuable works that have been published in the field of  XAI for SE. Interesting ideas have been suggested, and an encouraging growing trend has started. However, despite all of these efforts, there is a long way ahead. ML models are improving their performance and growing more complex gradually and yet there is a noticeable lack of interest from software practitioners in using them in the real world, partly due to the lack of transparency and trust issues.

As we mentioned in Section \ref{missing_areas} there are still unexplored areas such as software design and testing, and tasks in SE that already have taken advantage of ML models and yet have not been studied in terms of explainability. Code comment generation, code generation, vulnerability
detection, test case generation, bug classification, and program repair are some of the most important tasks that require more attention.

Additionally, if we consider the mentioned unexplored tasks, we can see a pattern of overlooking generation-based tasks (e.g. code generation, test case generation, and program repair). In our analysis, we discussed how classification problems are preferable for XAI due to their limited output space, but looking at recent advancements of AI models in generation-based tasks, we believe there is an undeniable need for XAI methods for those tasks.

A further note on this topic is these unexplored tasks and in general, all generation-based tasks is that their advancements rely heavily on deep neural networks. DNNs have been a great asset for researchers when it comes to searching in a huge space (i.e. vocabulary of words), thus they are ideal for this type of task. Code models and their success in code document generation, code translation, and code repair are good examples. However, they are incomparably more complex than classical ML models, so they are harder to explain. Nonetheless, this complexity is the exact reason that they require explanation.

Furthermore, by analyzing the XAI techniques used in SE models, we see that the generated explanations are mostly meant for developers and programmers and not other types of less-technical end users (e.g., managers, executives, and other stakeholders) as there are fewer high-level explanations (such as visualizations or natural language rules) offered. While in other areas such as Image processing or NLP, the explanations are more user-friendly. This problem is partly because of the type of data and nature of SE tasks, but still, a lack of effort in giving such visualizations is noticeable.

Finally, if we look at our selected studies, we can see a clear lack of evaluation consistency among them. Some studies just explain the ML models for the sake of explaining and pay no more interest in further analysis, evaluation, or justification for the provided explanations. Considering the fact that explanations are supposed to help the models' users, we believe when objective evaluation is not possible or sufficient, trying human evaluation by experts, is a satisfactory solution.

\section{Related Works}

There have been lots of surveys and literature reviews addressing the XAI topic in general or for specific fields. For instance, a study surveys some traditional ML models such as SVM, linear, and tree-based models and notes a trade-off between their performance and reliability~\citep{dovsilovic2018explainable}. They also mention the ambiguity and confusion among works addressing XAI in terms of taxonomy and definitions.

In an attempt to standardize the efforts in XAI, some studies has tried to clarify some of the definitions and terminology and insisted in distinguishing the terms ``interpretability'' and ``explainability'' as different terms with the first only being one of the features of the latter~\citep{gilpin2018explaining}. They also claim that different approaches to solve the problem of black-box models, usually are unilateral and fail to address different aspects of explainability. A similar approach is taken by others that offer a more updated insight and also specify challenges and opportunities ahead\cite{arrieta2020explainable}.

Besides works that surveyed or reviewed the literature on XAI, in general, there are valuable attempts to particularly review the state of XAI in specific fields or models. As expected, medical applications that have taken great advantage of ML models, in recent years, are also among the most critical and sensitive domains. In a research, more than 200 papers that have used XAI in deep learning models for medical image processing are analyzed, which itself shows the high interest of researchers in the field~\citep{van2022explainable}. They also provide some suggestions for future works including the need for medical experts to actively participate in designing interpretation methods.

Natural language processing is also one of the areas that benefitted from surveys and reviews to examine the state of XAI in the literature. For instance, A survey examines 50 papers that are related to XAI in NLP and categorizes and analyzes them according to different aspects like locality and globality or being self-explaining or post-hoc~\citep{danilevsky2020survey}. Furthermore, they note a debate among researchers about the very existence of a trade-off between explainability and performance. Similar to many other surveys in other fields, they also mention the lack of clear terminology and understanding of XAI methods and provide insightful guidelines.

There are also reviews or surveys that instead of tasks or fields of study focus on different types of data or models. Among these works, there are studies that focus on time-series data specifically~\citep{rojat2021explainable}, or analyze the advances in XAI for tabular data~\citep{sahakyan2021explainable}. There are other researches that study different XAI methods for deep neural networks~\citep{montavon2018methods} or specifically reviews XAI in deep reinforcement learning models~\citep{heuillet2021explainability}. 

Software engineering is also a field that has benefited from ML models for a long time and SLRs have been conducted to analyze the literature for ML or DL techniques application in SE~\citep{yang2022survey, watson2022systematic}. However, to the best of our knowledge, this is the first work offering a systematic literature review for XAI in software engineering.

\section{Conclusion}

%In this study, we systematically examine and analyze XAI methods employed in the field of software engineering and their different applications. After conducting a comprehensive search of all of the published studies in five libraries without any time restriction, more than 800 papers were found. After multiple rounds of automatic and manual short-listing and removing all unrelated papers, we came up with 24 relevant published papers in XAI4SE and chose them as our final selected studies.

%In RQ1, we focused on different software engineering areas and tasks that have leveraged XAI methods and offered some level of explanation, and also investigated the objectives of the XAI in those studies. In RQ2, we explored different XAI methods and techniques that were used in these studies and the type of explanation these models could provide for their users. In RQ3, we focused on the usefulness of XAI methods and how they were able to achieve their claimed objectives. We also were interested in different evaluation methods and metrics utilized in the selected studies for evaluating the XAI method or the generated explanation.

To the best of our knowledge, this is the first study to systematically review XAI in the domain of SE. By reviewing 24 relevant papers (out of over 800 automatically selected ones) and analyzing the collected data, we report the current state of research in this area and identify challenges and road ahead for future research.  

In particular, the review shows that among different stages of SDLC, QA and maintenance have been the most popular among the XAI researchers, and this popularity is hugely focused on the defect prediction task. On the other hand generation-based tasks (e.g. code generation, test case generation, and program repair) which are popular in the ML4SE community are rarely studied in the XAI4SE literature. 

The data also shows that the number one impact among all targeted objectives of applying XAI to ML4SE models has been improving the original ML model.

We also observe that self-explainable ML models, such as classic random forest, decision tree, regression models, and naive Bayes are among the most used XAI techniques and post-hoc technique are less used in the community, so far. 

In terms of the type of explanations, XAI4SE has been mainly focused on ML developers as the end user and provides low-level debugging type explanations such as ``feature summary statistic'' (e.g., Tables showing some extracted key features and their distribution~\cite{4}) and ``model internals'' (e.g., self-attention scores of Transformers~\cite{1} and backtracking CNN layers of a NN~\cite{23}) , and less on higher-level natural language-based explanations or visualization.

Finally, a lack of standard ways for evaluating XAI methods was clear among the studies, which has led to many different project-specific evaluation metrics, as well as human subject assessments.

 \bibliographystyle{elsarticle-num} 
 \bibliography{sample-base}

%% The Appendices part is started with the command \appendix;
%% appendix sections are then done as normal sections
\appendix
\newpage
\onecolumn

\renewcommand*{\arraystretch}{1.3}
\begin{longtable}{p{0.05\textwidth}p{0.3\textwidth}p{0.2  \textwidth}p{0.3  \textwidth}p{0.1 \textwidth}}
\caption{List of the selected 24 papers. Conference: [C], Journal: [J] }
\label{tab:final_list}
\\
Ref &
  Title &
  Authors &
  Publication Venue &
  Publication Year \\ \hline
\citep{5} &
  Investigating Explainability of Generative AI for Code through Scenario-Based Design &
  J. Sun, Q. V. Liao, M. Muller, M. Agarwal, S. Houde, K. Talamadupula, J.D. Weisz &
  Intelligent User Interfaces (IUI) {[}C{]} &
  2022 \\
\citep{28} &
  Predicting Defective Lines Using a Model-Agnostic Technique &
  S. Wattanakriengkrai, P. Thongtanunam, C. Tantithamthavorn, H. Hata, K. Matsumoto &
  Transactions on Software Engineering {[}J{]} &
  2022 \\
\citep{18} &
  JITLine: A Simpler, Better, Faster, Finer-grained Just-In-Time Defect Prediction &
  C. Pornprasit, C. Tantithamthavorn &
  Mining Software Repositories (MSR) {[}C{]} &
  2021 \\
\citep{25} &
  Just-in-Time Defect Prediction Technology based on Interpretability Technology &
  W. Zheng, T. Shen, X. Chen &
  Dependable Systems and their Applications (DSA) {[}C{]} &
  2021 \\
\citep{19} &
  Practitioners’ Perceptions of the Goals and Visual Explanations of Defect Prediction Models &
  J. Jiarpakdee, C. Tantithamthavorn, J. Grundy &
  Mining Software Repositories (MSR) {[}C{]} &
  2021 \\
\citep{15} &
  PyExplainer: Explaining the Predictions of Just-In-Time Defect Models &
  C. Pornprasit, C. Tantithamthavorn; J. Jiarpakdee, M. Fu, P. Thongtanunam &
  Automated Software Engineering (ASE) {[}C{]} &
  2021 \\
\citep{3} &
  Towards Reliable AI for Source Code Understanding &
  S. Suneja; Y. Zheng; Y. Zhuang; J. Laredo; A. Morari &
  Symposium on Cloud Computing {[}C{]} &
  2021 \\
\citep{10} &
  Understanding Neural Code Intelligence through Program Simplification &
  M. R. I. Rabin, V. J. Hellendoorn, M. A. Alipour &
  European Software Engineering Conference and Symposium on the Foundations of Software Engineering (ESEC/FSE) {[}C{]} &
  2021 \\
\citep{31} &
  Application of machine learning techniques to the flexible assessment and improvement of requirements quality &
  V. Moreno, G. Génova, E. Parra, A. Fraga &
  Software Quality {[}J{]} &
  2020 \\
\citep{23} &
  Deep Learning Based Valid Bug Reports Determination and Explanation &
  J. He, L. Xu, Y. Fan, Z. Xu, M. Yan, Y. Lei &
  International Symposium on Software Reliability Engineering (ISSRE) {[}C{]} &
  2020 \\
\citep{34} &
  Recognizing lines of code violating company-specific coding guidelines using machine learning: A Method and Its Evaluation &
  M. Ochodek, R. Hebig, W. Meding, G. Frost, M. Staron &
  Empirical Software Engineering {[}J{]} &
  2020 \\
\citep{21} &
  The Impact of Class Rebalancing Techniques on the Performance and Interpretation of Defect Prediction Models &
  C. Tantithamthavorn, A. E. Hassan, K. Matsumoto &
  Transactions on Software Engineering {[}J{]} &
  2020 \\
\citep{1} &
  An Explainable Deep Model for Defect Prediction &
  J. Humphreys, H. K. Dam &
  Workshop on Realizing Artificial Intelligence Synergies in Software Engineering (RAISE) {[}C{]} &
  2019 \\
\citep{4} &
  Neural Network-Based Detection of Self-Admitted Technical Debt: From Performance to Explainability &
  X. Ren, Z. Xing, X. Xia, D. Lo, X. Wang, J. Grundy. &
  Transactions on Software Engineering and Methodology (TOSEM) {[}J{]} &
  2019 \\
\citep{12} &
  Requirements Classification with Interpretable Machine Learning and Dependency Parsing &
  F. Dalpiaz, D. Dell'Anna, F. B. Aydemir, S. Çevikol &
  Requirements Engineering (RE) {[}C{]} &
  2019 \\
\citep{9} &
  Towards a More Reliable Interpretation of Defect Models &
  J. Jiarpakdee &
  International Conference on Software Engineering: Companion (ICSE-Companion) {[}C{]} &
  2019 \\
\citep{6} &
  Explainable Software Analytics &
  H. K. Dam, T. Tran, A. Ghose &
  International Conference on Software Engineering: New Ideas and Emerging Results (ICSE-NIER) {[}C{]} &
  2018 \\
\citep{14} &
  Superposed Naive Bayes for Accurate and Interpretable Prediction &
  T. Mori &
  International Conference on Machine Learning and Applications (ICMLA) {[}C{]} &
  2015 \\
\citep{16} &
  A novel method for software defect prediction: Hybrid of FCM and random forest &
  T.P. Pushphavathi, V. Suma, V. Ramaswamy &
  International Conference on Electronics and Communication Systems (ICECS) {[}C{]} &
  2014 \\
\citep{32} &
  Software defect prediction using Bayesian networks &
  A. Okutan, O. T. Yıldız &
  Empirical Software Engineering {[}J{]} &
  2014 \\
\citep{13} &
  Towards Interpretable Defect-Prone Component Analysis Using Genetic Fuzzy Systems &
  T. Diamantopoulos, A. Symeonidis &
  Workshop on Realizing Artificial Intelligence Synergies in Software Engineering (RAISE) {[}C{]} &
  2014 \\
\citep{7} &
  Software Effort Prediction: A Hyper-Heuristic Decision-Tree Based Approach &
  M. P. Basgalupp, R. C. Barros, T. S. da Silva, A. C. P. L. F. de Carvalho &
  Symposium on Applied Computing {[}C{]} &
  2013 \\
\citep{27} &
  Machine-Learning Models for Software Quality: A Compromise between Performance and Intelligibility &
  H. Lounis, T. F. Gayed, M. Boukadoum &
  International Conference on Tools with Artificial Intelligence {[}C{]} &
  2011 \\
\citep{33} &
  Evaluation of preliminary data analysis framework in software cost estimation based on ISBSG R9 Data &
  Q. Liu, W. Z. Qin, R. Mintram, M. Ross &
  Software Quality {[}J{]} &
  2008
\end{longtable}

\twocolumn

%% else use the following coding to input the bibitems directly in the
%% TeX file.

% \begin{thebibliography}{00}

% %% \bibitem{label}
% %% Text of bibliographic item

% \bibitem{}

% \end{thebibliography}
\end{document}